\documentstyle[aps,floats,manuscript,epsfig]{revtex}
\newcommand{\beq}{\begin{equation}}
\newcommand{\eeq}{\end{equation}}
\newcommand{\ba}{\begin{array}}
\newcommand{\ea}{\end{array}}
\begin{document}
\title{Electron-neutrino survival probability from 
solar-neutrino data}

\author{V.~Berezinsky\cite{email1}}
\address{INFN, Laboratori Nazionali del Gran Sasso, I-67010 Assergi (AQ), 
Italy, \\
and Institute for Nuclear Research, Moscow, Russia}
\author{M.~Lissia\cite{email2}}
\address{Istituto Nazionale di Fisica Nucleare, Sezione di Cagliari and 
      Dipartimento di Fisica dell'Universit\`a di Cagliari,
      I-09042 Monserrato (CA), Italy}
\maketitle
\begin{abstract}
With SNO data \cite{SNO} on electron-neutrino flux from the sun, it is 
possible to derive the $\nu_e$ survival probability $P_{ee}(E)$ 
from  existing experimental data of Super-Kamiokande, gallium experiments and 
Homestake. The combined data of SNO and Super-Kamiokande provide boron 
$\nu_e$ flux 
and the total flux of all active boron neutrinos, giving  thus $P_{ee}(E)$ 
for boron neutrinos. The Homestake detector, after subtraction of the signal
from boron neutrinos, gives the flux of Be+CNO neutrinos, and $P_{ee}$ 
for the corresponding energy interval, if the produced flux is taken from 
the Standard Solar Model (SSM).  
Gallium detectors, GALLEX, SAGE and GNO, detect additionally
$pp$-neutrinos. The $pp$ flux can be calculated subtracting from the gallium
signal the rate due to boron, beryllium and CNO neutrinos. The ratio of 
the measured $pp$-neutrino flux to that predicted by the SSM gives the survival
probability for $pp$-neutrinos. 
Comparison with theoretical survival 
probabilities shows that the best (among known models) fit is given by 
LMA and LOW solutions.    
\end{abstract}
\section*{}
The recent measurement of boron $\nu_e$ flux by SNO \cite{SNO},
combined with Super-Kamiokande data \cite{SK}, gives strong evidence for
neutrino oscillations \cite{SNO}. It is based on the fact that
the $\nu_e$ flux measured by SNO through the charged current (CC) interaction
$\nu_e+d \rightarrow e+p+p$~ induces in the Super-Kamiokande detector less 
electrons than observed. The excess of the electrons can be produced
only by active neutrinos of other flavors: $\nu_{\mu}$ and $\nu_{\tau}$.
This flux of active neutrinos is thus determined and found to be
$3.3 \sigma$ above the zero value. Such analysis was
suggested earlier in Ref.~\cite{Vill}, and recently it was further
developed in Ref.\cite{Lisi} (for other recent analysis of SNO data see  
\cite{Giunti,Bahcall}).

In this Letter we shall demonstrate that the new results obtained by
SNO allow us to derive the survival probability for boron, 
beryllium and $pp$ electron neutrinos (for general analysis 
see \cite{Barger} and for calculation of survival probability for
boron neutrinos \cite{Lisi}).

The probability of electron neutrinos to survive on the way from the
production point inside the Sun to the detection site on
the Earth is referred to as survival probability, $P_{ee}$. In case of
oscillations, $1-P_{ee}$ is the probability of electron-neutrino 
conversion into neutrinos of other flavors.

The flux of $^8$B electron neutrinos determined by SNO via CC-events
is $\Phi_{\nu_e} = 1.75 \pm 0.148 \cdot 10^6$~cm$^{-2}$s$^{-1}$ \cite{SNO},
where we used the upper value for systematic error and summed errors 
quadratically. SK measurements provide the flux 
$\Phi_{\nu}=\Phi_{\nu_e}+0.154(\Phi_{\nu_{\mu}}+\Phi_{\nu_{\tau}})$, 
where 0.154 is a ratio of cross-sections 
$\sigma_{\nu_{\mu} e}/\sigma_{\nu_e e}$. The comparison of these two
fluxes allows us to find $\Phi_{\nu_{\mu}}+\Phi_{\nu_{\tau}}$ and 
$\Phi_{\rm tot}=\Phi_{\nu_e}+\Phi_{\nu_{\mu}}+\Phi_{\nu_{\tau}}$, which is
equal to $5.44 \pm 0.99\times 10^6$~cm$^{-2}$s$^{-1}$\cite{SNO}. Thus the
survival probability for $^8$B electron neutrinos  can be found as 
$P_{ee}=\Phi_{\nu_e}/\Phi_{\rm tot}=0.32 \pm 0.065$. Note that this 
derivation of $P_{ee}$ does not depend on the SSM flux. The partial 
oscillation to sterile neutrinos, $\nu_e \rightarrow \nu_s$,
diminishes further $P_{ee}$. This possibility is somewhat disfavored 
by the observation that $\Phi_{\rm tot}$ is close to prediction of 
the SSM \cite{SNO}.

The error in the value of $P_{ee}$
indicated above  is calculated by adding all errors in quadrature and 
it needs further discussion. 
The uncertainties of $\Phi_{\nu_e}$ and $\Phi_{\rm tot}$ are
correlated and therefore 
the usual interpretation of the indicated error is allowed only in the limit 
$\Phi_{\nu_e}\ll \Phi_{\rm tot}$. In fact, the $2\sigma$- and 
$3\sigma$-equivalent intervals are asymmetric and greater than 
$1\sigma$ error 0.065
multiplied by factor of 2 and 3, respectively. In particular it can be
shown that $1-P_{ee}$ interval has real 3.3$\sigma$ deflection from zero
value. 

The value of $P_{ee}$ is plotted in
Figs. 1--3. The calculated value refers to the whole energy interval 
of boron neutrinos measured in Super-Kamiokande: in Figs.1--3
the horizontal error bars show the width of this interval.
The value of $P_{ee}$ is plotted in the middle of this interval and 
looks asymmetric in logarithmic scale.
As already mentioned above, one should not interpret the
many standard deviations which separates $P_{ee}$ from unity in the
usual way: the probability that $P_{ee}=1$ corresponds to about
$3.3\sigma$ and not to  $\approx 10\sigma$  as it might appear from
the figure.
\vspace{-5mm}
\begin{figure}[hp]
\begin{center}
\epsfig{bbllx=90pt,bblly=250pt,bburx=520pt,bbury=670pt,%
file=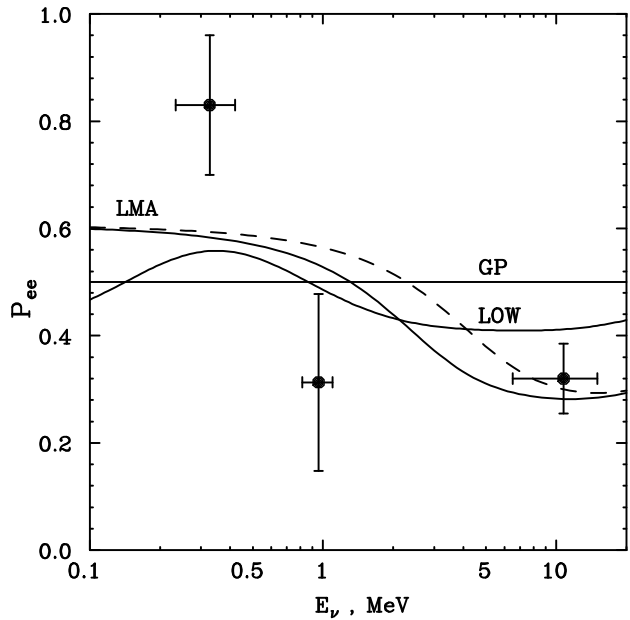,height=9cm}
\end{center}
\vspace{-5mm}
\caption[a]{Survival probability of electron neutrinos as a function
of energy. Data points are extracted from
the gallium, chlorine and boron-neutrino signals. The horizontal error
bars give the energy windows of each datum. For the 
interpretation of vertical error bars see text. The three solid curves show
the theoretical survival probabilities for the Gribov-Pontecorvo (GP) solution,
the LOW MSW solution (LOW) and for the large mixing angle MSW (LMA) solution.
The solid LMA curve corresponds to neutrino production point at the
peak of the boron/beryllium production zone; the dashed curve -- 
at the peak of the $pp$ production zone. 
\label{fig1}
} 
\end{figure}

The Homestake detector is sensitive to the boron, beryllium and CNO 
$\nu_e$-neutrinos. Subtracting the contribution of
the measured flux of boron $\nu_e$-neutrinos with the standard
spectrum to the chlorine
detector, we determine the contribution of Be+CNO 
neutrinos to the signal. For the production fluxes we use that of 
the SSM \cite{BP00}: note that the
prediction for the Be neutrino flux is reliable and the contribution of CNO 
neutrinos is small. In this case the extracted survival
probability $P_{ee}$ is not affected by possibility of 
$\nu_e \rightarrow \nu_s$ oscillation.
The survival probability is plotted in Figs.1--3
with errors calculated taking into account uncertainties in
fluxes together with statistical and systematic errors of detection. 
The horizontal error bar refers to the energy interval of Be+CNO neutrinos.
The error is strongly anticorrelated with the error in the boron flux:
a higher (lower) boron flux implies a lower (higher) beryllium flux.

The detected $\nu_e$-neutrino flux of pp-cycle is found from gallium
experiments (GALLEX, SAGE and GNO), subtracting the contributions from
boron, beryllium  and CNO neutrinos found from the Kamiokande, SNO and
gallium experiments. The production flux is taken from SSM
calculations \cite{BP00}.
The calculated $pp$ flux is robust and agrees with the flux found
independently from solar-luminosity sum rule. The survival probability
is plotted in Figs. 1-3. The horizontal error bar refers to the pp-neutrino
energy interval. The survival probability is not affected by
oscillation to sterile neutrinos.
\begin{figure}[h]
\begin{center}
\epsfig{bbllx=90pt,bblly=250pt,bburx=520pt,bbury=670pt,%
file=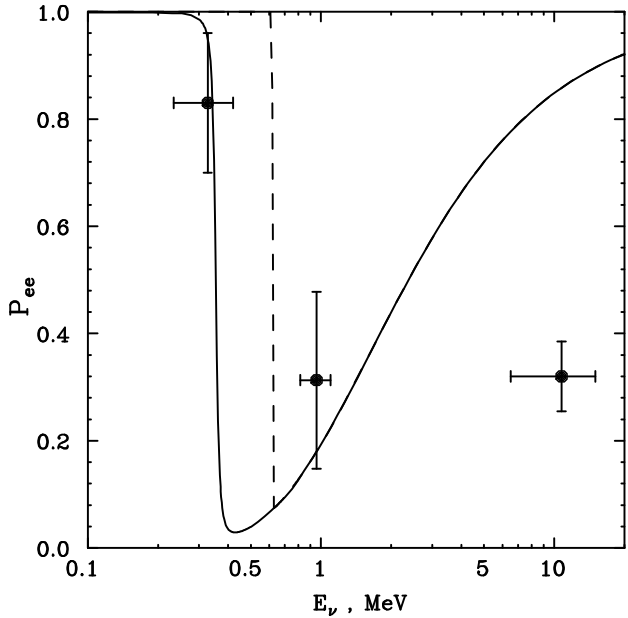,height=9cm}
\end{center}
\caption[b]{Survival probability of electron neutrinos as function of energy.
The solid curve shows the theoretical survival probability for the small
mixing angle (SMA) MSW solution for neutrino born at the
peak of the
boron/beryllium production region; the dashed curve -- at  the peak  
of the $pp$ production region.
\label{fig2}
} 
\end{figure}
We shall summarize now the assumptions involved in the calculations 
of survival probabilities at different energies. For boron
neutrinos we used only experimental data, neglecting possible 
oscillation to sterile neutrinos.
For the other two energy
intervals ($pp$- and Be-neutrinos) we used for total fluxes the 
production fluxes calculated in the SSM, 
but the calculations of these fluxes are reliable. In addition, we
assumed that the
measured suppression of high energy boron neutrinos
(roughly for $E>6$ MeV) is valid also for the lower part of the 
boron-neutrino spectrum.
Since in both cases only electron neutrinos produce the signal,
the extracted survival probability is not affected by oscillation to
sterile neutrinos.

In Fig.~1 the observed survival probabilities are compared with
the predictions of the LMA MSW, LOW MSW and
Gribov-Pontecorvo (GP) \cite{GP} solutions.
For LMA we use one of the best fit solutions from Ref.~\cite{BCC}:
$\Delta m^2 = 3.7\times 10^{-5}$~eV$^2$
and $\sin^2 2\theta=0.79$;
for LOW: $\Delta m^2 = 1.0\times 10^{-7}$~ eV$^2$, and 
$\sin^2 2\theta= 0.97$ \cite{BCC}, and for the Gribov-Pontecorvo solution
\cite{GP} $\sin^2 2\theta=1$. The LMA and LOW survival probabilities
shown are not averaged over the production points in the sun. The LOW curve
is practically independent of the production point. The LMA solid
curve is shown
for a production point in the boron/beryllium production zone and  
the dashed curve shows the survival probability for a neutrino produced at
the peak of the $pp$ region.  Averaging over the production region gives
a curve close to the solid one for boron and beryllium neutrinos, and a
curve between the solid and dashed one for $pp$ neutrinos (in fact
there is little difference between these curves in the energy region of 
$pp$ neutrinos).\\
Only for {\em illustration}, we present $\chi^2/d.o.f.$ values for
the fits of the data by different oscillation solutions: they are  
4/3, 5/3 and 10/3 for the LMA, LOW and GP solutions, respectively.
In fact, this $\chi^2$ analysis could be misleading because when the
survival probability is used as a variable, the probability distribution
is not Gaussian.
\begin{figure}[hp]
\begin{center}
\epsfig{bbllx=90pt,bblly=250pt,bburx=520pt,bbury=670pt,%
file=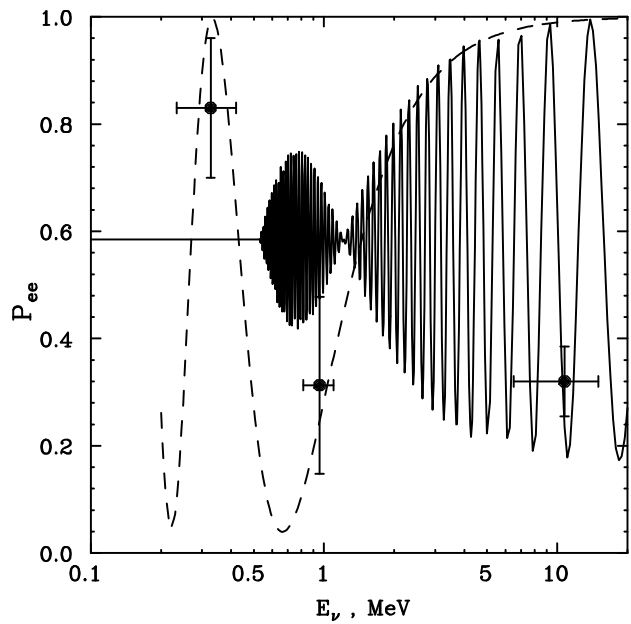,height=9cm}
\end{center}
\caption[c]{Survival probability of electron neutrinos as function of energy.
The solid curve shows the theoretical survival probability for the vacuum
oscillation solution; for energy below 0.53~MeV, only the average probability
is shown. The dashed curve is for the Just-So$^2$ solution. 
\label{fig3}
} 
\end{figure}

In Fig.~2 the observed survival probabilities are  compared with 
those calculated in SMA MSW. One of the best fit solutions, with spectral and 
temporal information included, is given by  
$\Delta m^2=4.6\times 10^{-6}$~ eV$^2$ and
$\sin^2 2\theta =1.36 \times 10^{-3}$.
The solid curve corresponds again
to the neutrino produced in the boron/beryllium production region, while the
dashed curve shows the low-energy part of the survival probability for the 
neutrino born in the
peak of $pp$ production region. The confidence level 
is characterised {\em illustratively} by $\chi^2/d.o.f. \approx 12/3$. 

In Fig.~3 the observed survival probabilities are compared with 
predictions of the Just-So ($\Delta m^2=4.6\times 10^{-10}$~eV$^2$ 
and  $\sin^2 2\theta=0.83$)
and Just-so$^2$ ($\Delta m^2=5.5\times 10^{-12}$~eV$^2$ and
$\sin^2 2\theta=0.96$) solutions. The qualitatively worse agreement can
be characterised by $\chi^2/d.o.f. \approx 11/3 $ and
$\chi^2/d.o.f. \approx 13/3$ for Just-So and Just-So$^2$, respectively.

We repeat again that the quantities $\chi^2$ calculated above 
have only an illustrative character: the obtained value 
of $\chi^2$ for each solution is not connected with probability in a 
usual way. This analysis indicates however the preferred solutions.

The preferred solutions are  LMA  and LOW which are characterised by 
the lowest values of $\chi^2$.

In principle the survival probabilities for boron neutrinos 
can be given for several energy bins. Since the observed spectrum is
well described by the SSM spectrum, the energy bin analysis will 
further decrease the probability of
SMA MSW, which produces a significant spectrum distortion in the 
boron energy window, and will favour the solutions with a flat 
suppression of boron neutrino spectrum, such as
the GP, LMA and LOW solutions. Day-night dependence will not change the
analysis in a significant way, since these solutions have little
day-night dependence in the boron high-energy window.

In conclusion, using the data of all solar-neutrino experiments 
we have derived the electron-neutrino survival probabilities. For boron 
neutrinos only experimental data are used, and the SSM flux is not
involved.
Survival probability decreases in presence of 
$\nu_e \rightarrow \nu_s$ oscillation.
For Be+CNO and $pp$ neutrinos the SSM calculations are used for production
(i.e total) fluxes. This is a plausible assumption, because beryllium and $pp$
neutrino fluxes are reliably calculated and CNO fluxes are small. 
Oscillation to sterile neutrinos does not affect the extracted survival  
probabilities for Be+CNO and $pp$ neutrinos.

LMA and LOW solutions give better fits, in comparison with other
solutions. It is also possible, in principle, to calculate for boron 
neutrinos the survival probabilities for several energy bins: the
likelihood of solutions with an energy independent suppression for $E>6$~MeV,
i.e. LOW, LMA and GP, will increase as compared with the ones that predict
an energy dependent suppression.

\end{document}